\documentclass[%
reprint,
superscriptaddress,
showpacs,
amsmath,amssymb,
aps,
prl,
longbibliography,
]{revtex4-2}

\usepackage{psfrag,graphicx,epsfig,color}
\usepackage{dcolumn}
\usepackage{bm}
\usepackage{natbib}
\usepackage{float}
\usepackage[usenames,dvipsnames,svgnames,table]{xcolor}
\usepackage{subfigure}
\usepackage{nicefrac}

\def\re    {Re_\lambda}
\def\uu {{\mathbf{u}}}

\def\ww {{\boldsymbol{\omega}}}

\definecolor{mygreen}{rgb}{0,0.7,0.}

\begin{document}

\title{
Universality of extreme 
events in turbulent flows
}

\author{Dhawal Buaria }
\email[]{dhawal.buaria@ttu.edu}
\affiliation{Department of Mechanical Engineering, Texas Tech University, Lubbock, TX 79409}
\affiliation{Max Planck Institute for Dynamics and Self-Organization, 37077 G\"ottingen, Germany}
\author{Alain Pumir}
\affiliation{Ecole Normale Superieure, Universite de Lyon 1 and CNRS, 69007 Lyon, France}
\affiliation{Max Planck Institute for Dynamics and Self-Organization, 37077 G\"ottingen, Germany}


%

%

\date{\today}

\begin{abstract}


The universality of small scales,
a cornerstone of turbulence,
has been nominally confirmed for low-order mean-field
statistics, such as the energy spectrum. 
However, small scales exhibit strong
intermittency, exemplified by formation of extreme events
which deviate anomalously from a mean-field description. 
Here, we investigate the universality of small scales by analyzing
extreme events of velocity gradients in different turbulent flows, 
viz. direct
numerical simulations (DNS) of homogeneous
isotropic turbulence, inhomogeneous channel flow, and 
laboratory measurements in a von Karman mixing tank.
We demonstrate that the scaling exponents of velocity gradient moments,
as function of Reynolds number ($Re$), are universal, in agreement with 
previous studies at lower $Re$, and further show that
even proportionality constants are universal when considering
one moment order as a function of another. 
Additionally, by comparing various unconditional and conditional 
statistics across different flows, we 
demonstrate that the structure of the velocity gradient tensor
is also universal. 
Overall, our findings provide compelling evidence that even
extreme events are universal, with profound implications
for turbulence theory and modeling. 


\end{abstract}

\maketitle



In turbulent flows, energy is injected at large scales,
and cascades across a wide range of intermediate scales down to very
small scales, where it is dissipated into heat. 
The dynamics at large-scales depend on the geometry and 
specifics of the flow considered, rendering them
inherently anisotropic and non-universal.  
However, as energy cascades to smaller scales, the influence 
of large-scales diminishes, leading to small scales
becoming increasingly isotropic and universal. 
This notion underpins the foundational principle
of small-scale universality, 
first proposed by Kolmogorov (1941) \cite{K41a}---henceforth 
K41---which is a cornerstone of turbulence theory and modeling.
It asserts that when the scale separation is sufficiently large
(or equivalently the flow Reynolds number $Re$ is large), the statistical 
properties of the small scales exhibit universal behavior, 
consistent across different turbulent flows.
K41 further hypothesizes that
the small scales are solely characterized by the fluid viscosity
$\nu$
and the mean dissipation rate $\langle \epsilon \rangle$, which captures
the net transfer of energy across the scales.

The most compelling support for small-scale universality 
comes from
the well-known $-5/3$ scaling of the energy spectrum
in the inertial range (where viscosity
can be neglected) and other similar results \cite{Grant:1962,Saddoughi:1994}.
In addition, studies have also investigated 
validity of local isotropy in the inertial
range and also for velocity gradients
\cite{kerr85, vanAtta91,Saddoughi:1994, BifProc}. 
However, while K41 has been generally successful in describing 
low-order statistics,
it is well-known that energy transfers in turbulence
are highly intermittent, with fluctuations
of dissipation rate, and velocity gradients in general,
exhibiting large deviations from the mean 
\cite{MS91, Jimenez93, Ishihara07, BPBY2019}. 
This phenomenon of 
intermittency, invalidates
K41's mean-field description \cite{Frisch95, Sreeni97}, 
and raises a natural question about the universality
of extreme events, which is the motivation
for this Letter. 

Despite its obvious importance, the universality of extreme events 
has received limited attention.
This gap arises primarily from challenges associated with
measuring the full velocity gradient tensor
in experiments \cite{Wallace09}.
Similarly,
direct numerical simulations (DNS) were
historically limited
to lower $Re$  due to their high computational cost \cite{mm98};
with the constraints being even more severe for resolving extreme events 
\cite{Paladin87, YS:05, PK+18, BPBY2019}.
Nevertheless, DNS studies at low $Re$ by \cite{Schumacher+14,Ham+12} 
have provided some support for universality 
by demonstrating identical scaling exponents 
of moments of dissipation-rate as function of $Re$. 
However, this only addresses part of the question
and does not fully capture the structural complexity of velocity gradient
tensor, which encompasses various non-trivial correlations between
strain-rate and vorticity \cite{Ashurst87, BBP2020, Tsi2009}.

In this Letter, we address the universality of extreme events
of velocity gradients,
by examining turbulence in three distinct flows:
DNS of homogeneous isotropic turbulence (HIT), DNS of plane
channel flow, and laboratory experiments
in a von K\'arm\'an mixing tank. 
These flows are all governed by the incompressible Navier-Stokes equations, 
differing only in large-scale geometry and forcing mechanisms. 
The DNS data utilized are at substantially higher $Re$ than earlier
studies, with the HIT runs corresponding to unprecedented small-scale resolution
\cite{BPB2020, BP2022}. 
Concurrently, the experimental data provides knowledge
of the full velocity gradient tensor \cite{Knutsen:2020}, 
providing structural information on extreme events which was not
available previously.

By analyzing various statistics and the structure of 
the gradient tensor: $A_{ij} \equiv \partial u_i / \partial x_j$
($i,j=1,2,3$),  we provide strong evidence in support of 
universality of extreme events. 
First, by considering velocity gradients moments, we show
that their scaling exponents in $Re$ are same in different
flows, which substantially extends the results of \cite{Schumacher+14}
to higher $Re$ and broader range of flow configurations.
We then further show that even the proportionality constants 
can be matched across different flows, by considering one 
moment-order as a function of another, 
akin to extended self-similarity \cite{Benzi+93}.
Beyond statistical moments, we also show that 
structural properties of extreme events,
as captured by various conditional statistics,
are quantitatively same across different flows, including
their $Re$-dependence.

\paragraph*{Data:}
We only briefly describe the data used here
since the three flows considered have been well studied
on their own (though some additional details are also
given in the Appendix A). 
We utilize
the Taylor-scale Reynolds number
$Re_\lambda \equiv u' \lambda /\nu$ for each flow, 
with $u'$ being the rms of velocity fluctuations, and 
$\lambda = u' / A' $ the
Taylor length scale, where $A'$ is the rms of 
longitudinal (diagonal) components of $A_{ij}$.
Note that $Re_\lambda \sim Re^{1/2}$ \cite{Frisch95}.
For HIT, the DNS data corresponds to several
recent works 
\cite{BBP2020, BPB2020, BS2020, BP2021, BPB2022, BS2022, BP2023}, 
with the Taylor scale Reynolds
number $Re_\lambda$ going from $140$ to $1300$,
on grids of up to $12288^3$ points.
For channel flow, we utilize 
the Johns Hopkins turbulent database \cite{Graham+15}, 
with the skin-friction Reynolds numbers $Re_\tau = 1000$  
and $Re_\tau = 5200$. 
The experimental data 
corresponds to that of 
\cite{Knutsen:2020}, with $\re \approx 200$.
While HIT is necessarily isotropic, it should be noted
that plane channel flow is strongly anisotropic near
the wall. In fact, it is known that the presence of a
mean-shear persistently violates local isotropy
\cite{Shen:2000,Pumir:2016}.
Consequently, 
for channel flow, statistics are 
obtained only in the outer region, in a slab
around the centerline, where the mean shear is
very weak; this is also consistent with the approach
of \cite{Schumacher+14}. The measurement volume
in the von K\'arm\'an flow is always at the center,
where large-scale mean-gradient, if any, are very small \cite{Knutsen:2020}.

\paragraph*{Results:}
Before considering extreme events, it is worth highlighting
the universal aspects of velocity gradients for the mean field
itself. To that end, we consider the strain
tensor $S_{ij} = (A_{ij} + A_{ji})/2$,
and the vorticity vector $\omega_i = \varepsilon_{ijk} A_{jk}$,
(where $\varepsilon_{ijk}$ is the Levi-Civita symbol).
The strain tensor can further be decomposed
into an orthonormal basis, identified by
three eigenvectors $\mathbf{e}_i$
and corresponding eigenvalues $\lambda_i$,
such that $\lambda_1 \geq \lambda_2 \geq \lambda_3$;
Incompressibility imposes $\sum_i \lambda_i = 0$,
implying $\lambda_1>0$ and $\lambda_3<0$.
A well known universal aspect of turbulence
is that $\lambda_2$ is positive on average, 
which is connected to energy cascade process \cite{Betchov56},
and vorticity preferentially aligns with $\mathbf{e}_2$
\cite{Ashurst87, BBP2020}.

\begin{figure}
\centering
\includegraphics[width=0.44\textwidth]{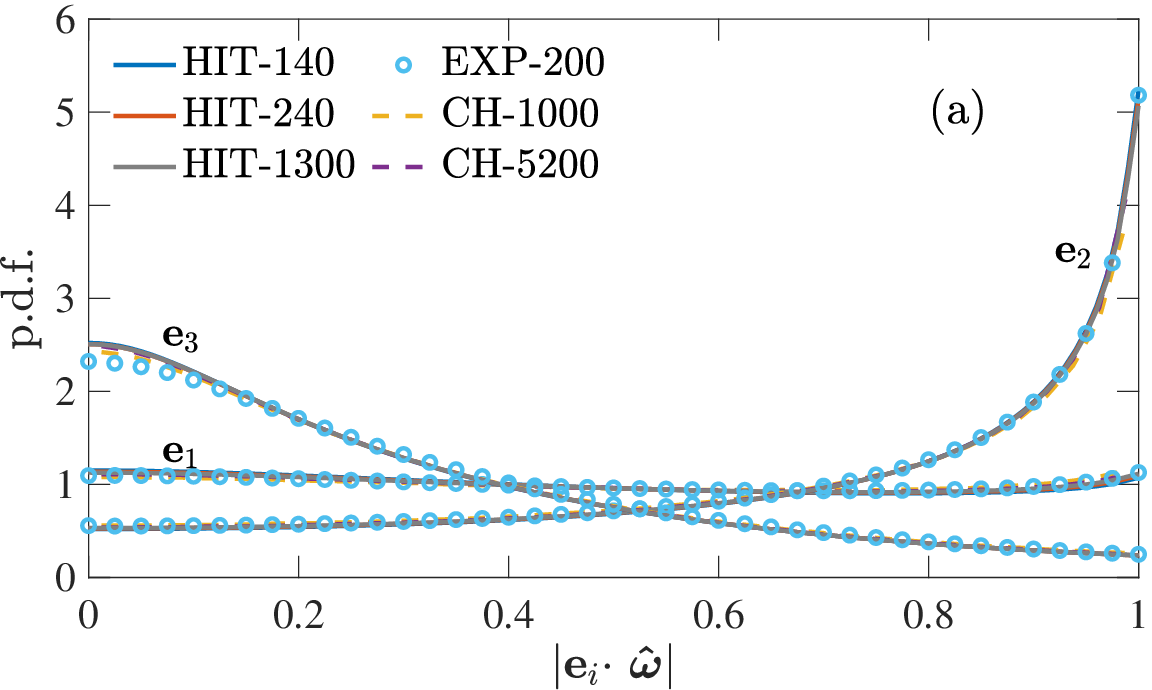} \\
\includegraphics[width=0.44\textwidth]{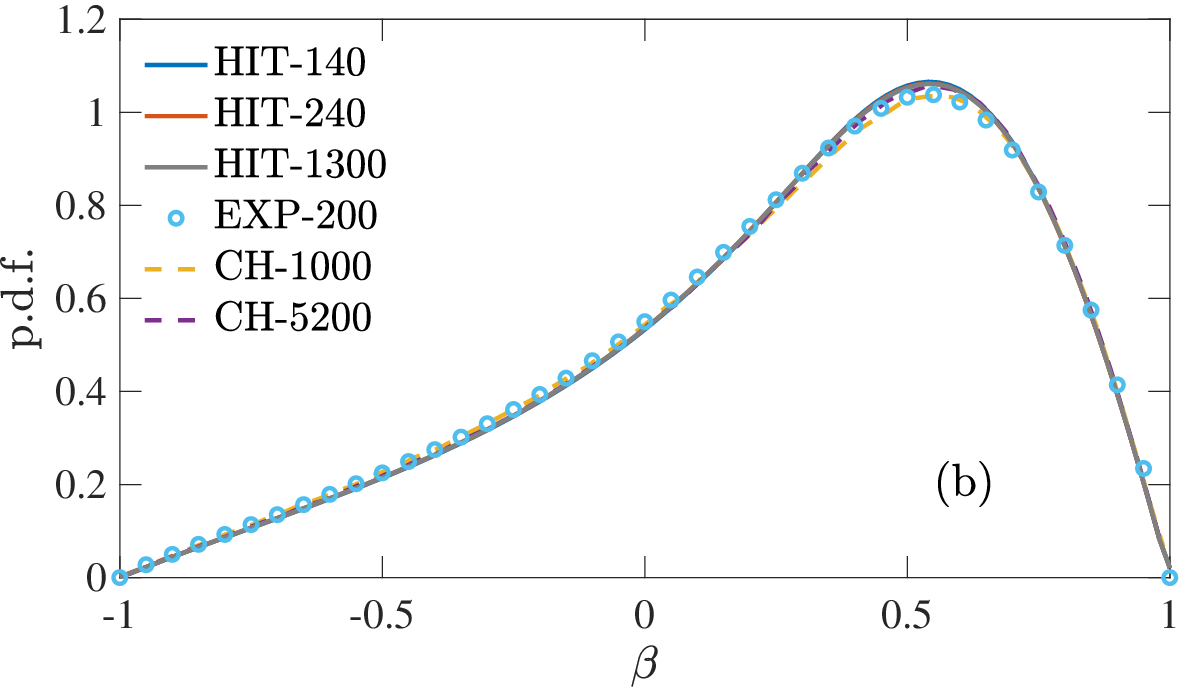} 
\caption{
Probability density functions (PDFs) of 
(a) alignment cosines between vorticity $\hat{\ww}$ and strain eigenvectors
($\mathbf{e}_i$), and (b) $\beta  = \sqrt{6} \lambda_2 / 
(\lambda_1^2 + \lambda_3^2 + \lambda_3^2)^{1/2}$. 
}
\label{fig:mean}
\end{figure}

Figure~\ref{fig:mean}a shows the probability density functions (PDFs) of the 
alignment cosines between vorticity and the strain eigenvectors: 
$|\mathbf{e}_i \cdot \hat{\ww}|$, where $\hat{\ww} = \ww / |\ww|$, for 
all the available data.
It can be observed that the superposition between curves is nearly
perfect. While ref.~\cite{BBP2020} already demonstrated 
$Re$-independence of these PDFs for HIT, 
our results here further demonstrate
universality of these PDFs across different flows (at all $Re$).
A similar conclusion is drawn from Fig.~\ref{fig:mean}b, which shows the PDF 
of $\beta  = \sqrt{6} \lambda_2 / 
(\lambda_1^2 + \lambda_3^2 + \lambda_3^2)^{1/2}$, 
which provides a relative measure of $\lambda_2$ 
with respect to the overall strain magnitude, with the constraint $|\beta| \leq1$.

K41 hypothesizes that all small-scale statistics, including those
of velocity gradients, are universal once rescaled by 
Kolmogorov length and time scales, respectively 
\begin{align}
\eta_K = (\nu^3 / \langle \epsilon \rangle)^{1/4} \ , \ \ \ 
\tau_K = (\nu / \langle \epsilon \rangle)^{1/2} \ .     
\end{align}
We focus on the non-dimensional moments of longitudinal (diagonal) 
components of $A_{ij}$,  i.e., 
\begin{align}
M_n \equiv \langle A_{\alpha 
\alpha }^n \rangle \tau_K^n  \ , 
\end{align}
for $\alpha=1,2,3$ and repeated
$\alpha$ not implying summation.
While K41 postulates that $M_n$ are constants, independent of $Re_\lambda$,
intermittency and extreme events lead to the following
dependence:
\begin{align}
M_n = c_n \re^{\xi_n} \ , \ \ \ \ \text{for} \ \ \ \re \gg 1 .
\label{eq:mn}
\end{align}
Since local isotropy gives 
$\langle \epsilon \rangle = 15 \nu \langle A_{\alpha \alpha}^2 \rangle$ 
\cite{Frisch95},
it readily follows that $M_2 = c_2 = 1/15$ and $\xi_2 = 0$.
but for $n \geq 3$, $\xi_n > 0$, 
and additionally, the constants $c_n$ are 
known to be flow-dependent \cite{Frisch95, Sreeni97, Schumacher+14}. 

\begin{figure}
\centering
\includegraphics[width=0.48\textwidth]{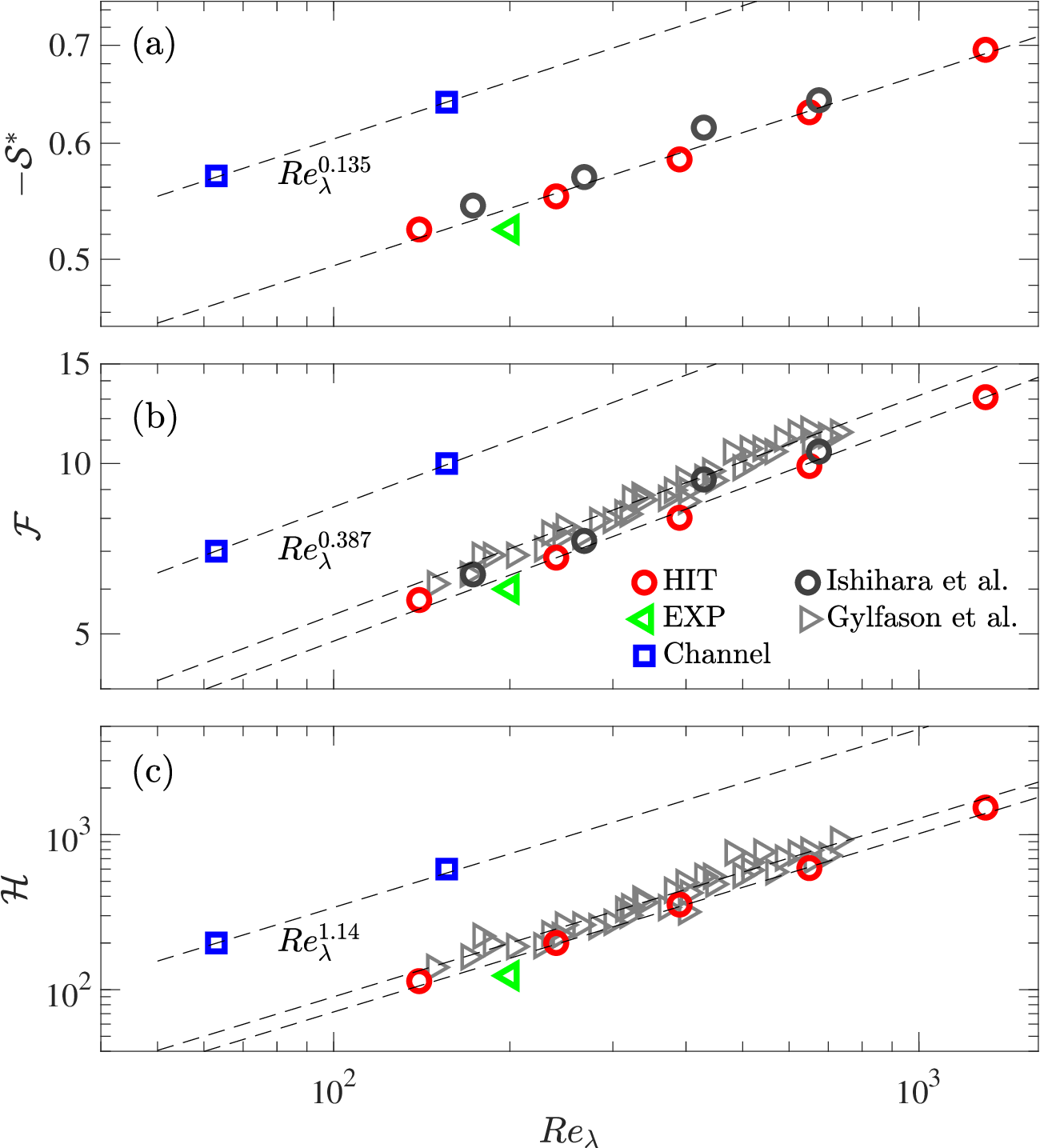} 
\caption{
(a) Skewness, (b) flatness, and (c) hyper-flatness
of longitudinal velocity gradients as a function of
$\re$ for different cases. Older DNS and experimental
data are used from Ishihara et al. \cite{Ishihara07} and 
Gylfason et al. \cite{gylfason2004},
respectively. All panels share the same legend.
}
\label{fig:mom1}
\end{figure}

To assess universality, we have extracted 
$M_n$ (up to $n=6$) from different flows. They are tabulated
in Table~\ref{tab:runs} in the Appendix B, 
together with some other details; we
reiterate the most important observations here.
It is observed that local
isotropy is valid for even moments of $A_{\alpha 
\alpha}$, i.e., $M_n$ are essentially independent of $\alpha$;
however, some persistent anisotropy is observed for the third
moment or the skewness
$\mathcal{S} = M_3/M_2^{3/2}$ 
(and also higher order odd moments, which are not shown).
As discussed in Appendix B, we instead consider an alternative
skewness 
$\mathcal{S^*} = - (6\sqrt{15}/7) \langle \omega_i \omega_j S_{ij} \rangle  \tau_K^3$
based on magnitude of vortex stretching, which is equal to $\mathcal{S}$ 
when local isotropy holds \cite{kerr85, BBP2020}, 
but otherwise provides an average measure 
of skewnesses in three directions.

Figure~\ref{fig:mom1}a-c respectively show the scaling of 
skewness $\mathcal{S}^*$, flatness $\mathcal{F} = M_4/M_2^2$, and 
hyper-flatness $\mathcal{H} = M_6/M_2^3$ 
as a function of $\re$.
Remarkably, we observe that the scaling exponents
are the same for all flows. While 
there is only one data point for newest experiments (EXP), 
we have also included previous experimental data 
from hot-wire measurements in grid turbulence \cite{gylfason2004},
and also earlier HIT DNS at lower $\re$ \cite{Ishihara07}.
Since only one component $A_{11}$
is available from hot-wire measurements \cite{gylfason2004}, 
we only consider even moments
for this data \footnote{as discussed in Appendix B, the third moments are
not strictly isotropic and hence using the skewness of just $A_{11}$
could be misleading}.
On the other hand, only up to fourth moments are available from \cite{Ishihara07}.
For all cases, the exponents $\xi_3 \approx 0.135$, $\xi_4 \approx 0.387$,
$\xi_6 = 1.14$  
are in excellent agreement with each other,
with earlier HIT studies
\cite{gylfason2004, Ishihara07, BS_PRL_2022, BS_PRF_2023},
and also with predictions from intermittency theories 
\cite{K62, Frisch95, YS:05, BS_PRL_2022}.
This confirms the universality of the scaling exponents $\xi_n$
in Eq.~\eqref{eq:mn}, with flow-dependent
prefactors $c_n$; 
in agreement with the observations of 
\cite{Schumacher+14},
who considered scaling of dissipation moments in HIT, channel flow,
and Rayleigh-Benard convection, albeit at lower
$Re$ than here
\footnote{see their Fig.4, where the data points
are shifted to identify the same scaling exponents,
but flow-dependent prefactors}.

\begin{figure}
\centering
\includegraphics[width=0.48\textwidth]{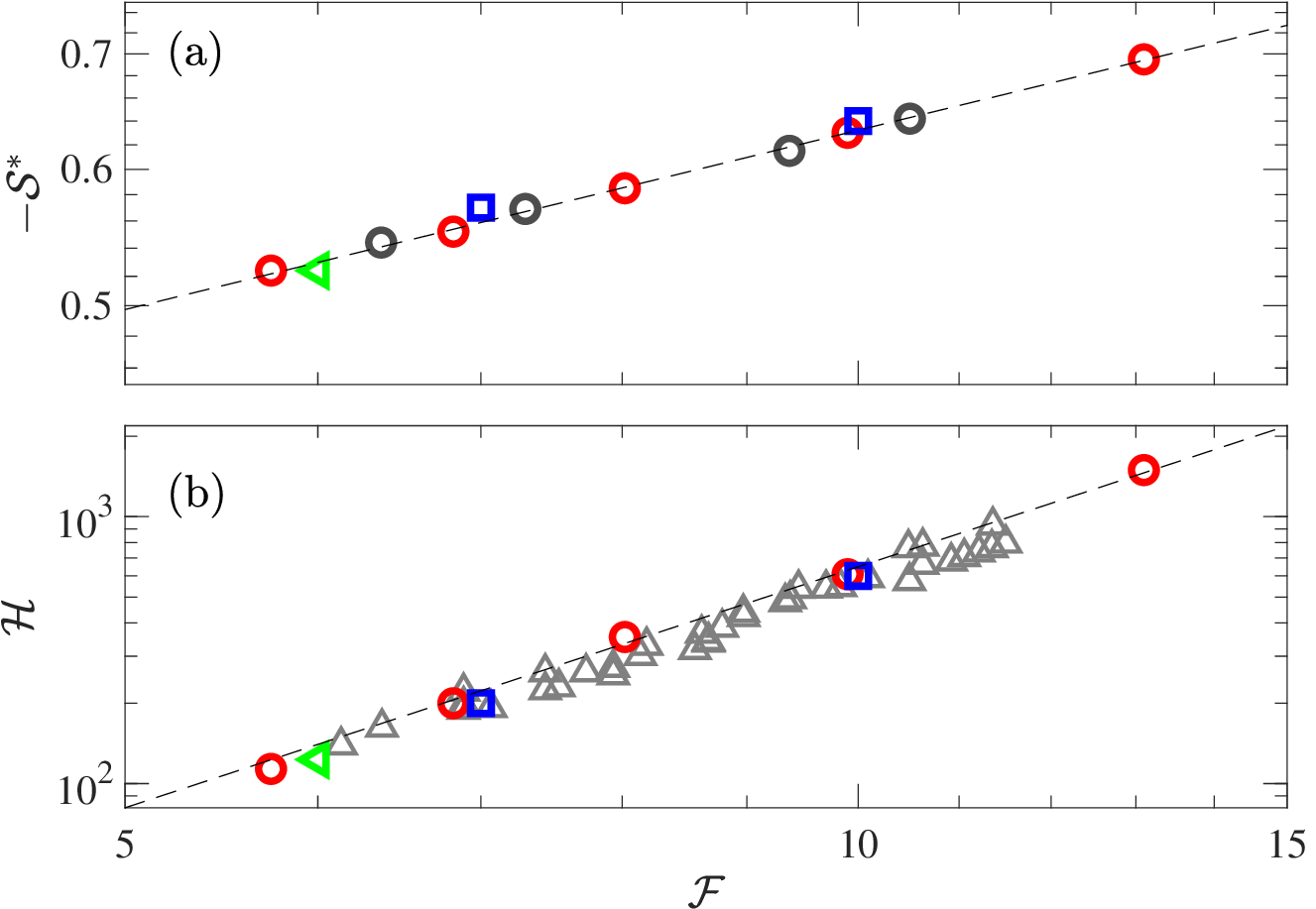} 
\caption{
(a) Skewness and (b) hyper-flatness as a function of flatness 
for different flows. Same legend as Fig.~\ref{fig:mom1} applies.
}
\label{fig:mom2}
\end{figure}

The above observation naturally leads to the question
about the role of Reynolds number when comparing different flows.
The ambiguity essentially arises from the fact that all Reynolds
number definitions use $u'$, which is the large-scale velocity
and hence, flow dependent. 
To resolve this ambiguity, we propose to use
a small-scale quantity to characterize the 
turbulence intensity as opposed to Reynolds number,
for instance, the skewness or flatness, 
which are both known to monotonically increase
with $\re$. 
Figure~\ref{fig:mom2}a shows the plot
of skewness vs. flatness, and  Fig.~\ref{fig:mom2}b shows
hyper-flatness vs. flatness, all taken from Fig.~\ref{fig:mom1},
which is in the spirit of extended 
self-similarity (ESS) \cite{Benzi+93}. 
Remarkably, we observe that
data from all different flows collapses on a single
curve (for both plots), which can be solely described by our HIT data.

The above result constitutes a substantially stronger evidence for universality
than previously suggested \cite{Schumacher+14}.
Essentially, using Eq.~\eqref{eq:mn},
the result in Fig.~\ref{fig:mom2}a can be
generally described
\begin{align}
M_n/M_2^{n/2} = K_n (M_4/M_2^2)^{\xi_n/\xi_4} \ ,   
\end{align}
where the prefactors $K_n$ are also universal
along with the exponents.
Note, the choice of using $M_4$ 
as the dependent variable is somewhat arbitrary,
and in principle, one can use any other moment. 
It is also worth mentioning here 
that recent Lagrangian results, 
see e.g. \cite{BS_PRL_2022, BS_PRL_2023}, also similarly provide support
for universality, but a more careful study is necessary, 
which we defer to future work.  


\begin{figure*}
\centering
\includegraphics[width=0.32\textwidth]{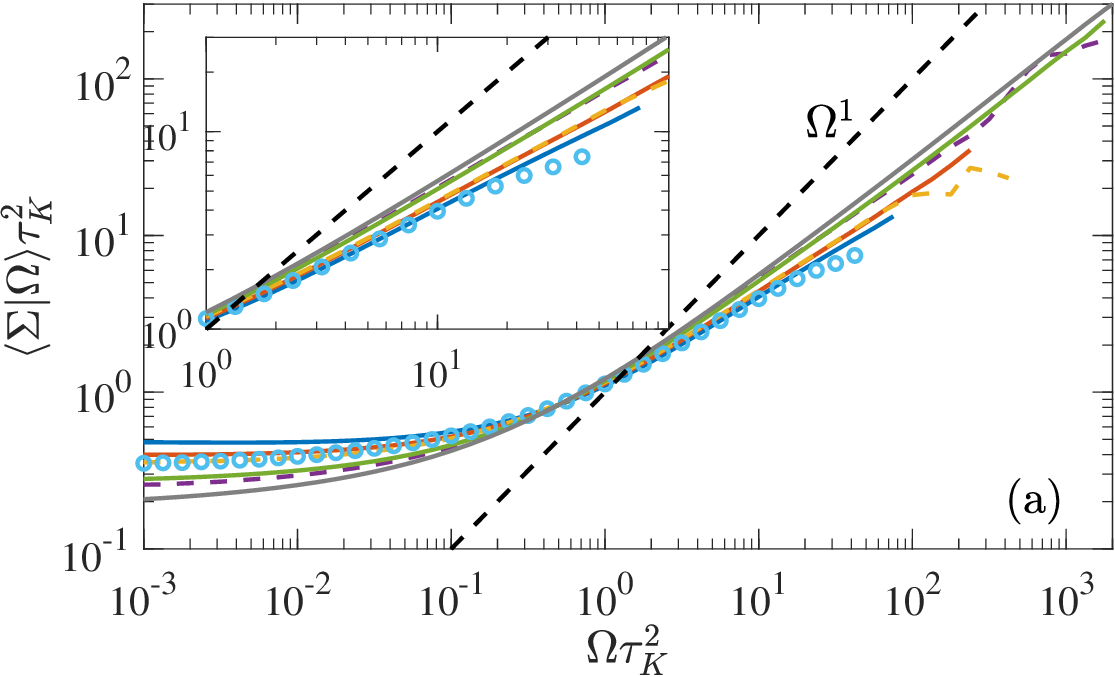} 
\includegraphics[width=0.32\textwidth]{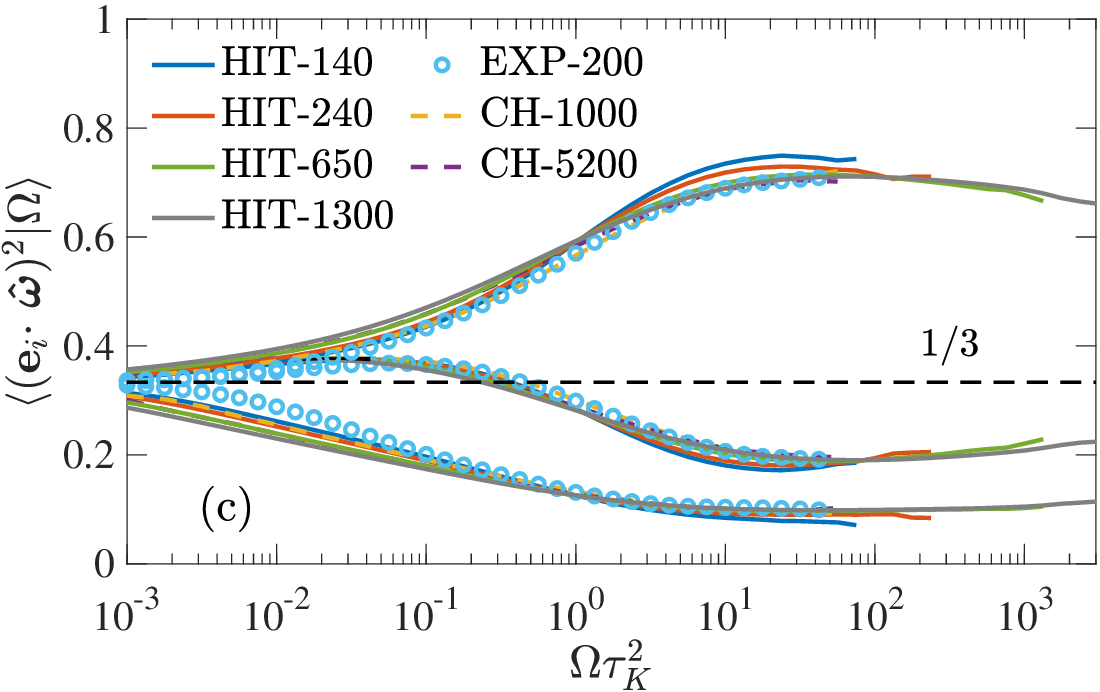} 
\includegraphics[width=0.32\textwidth]{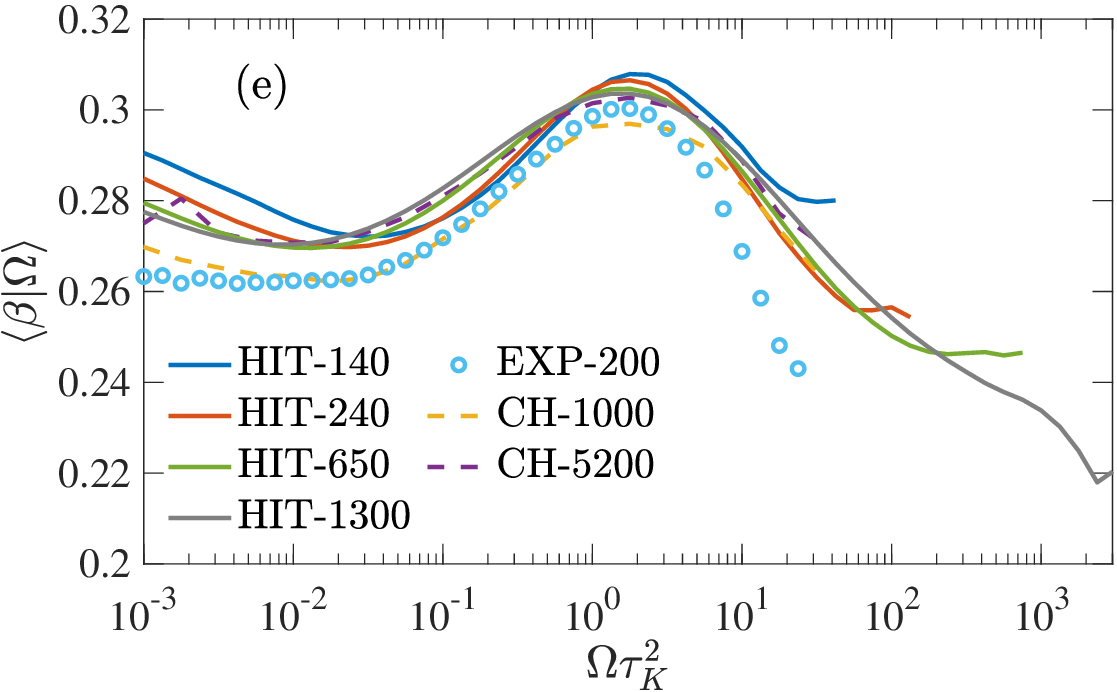} \\
\includegraphics[width=0.32\textwidth]{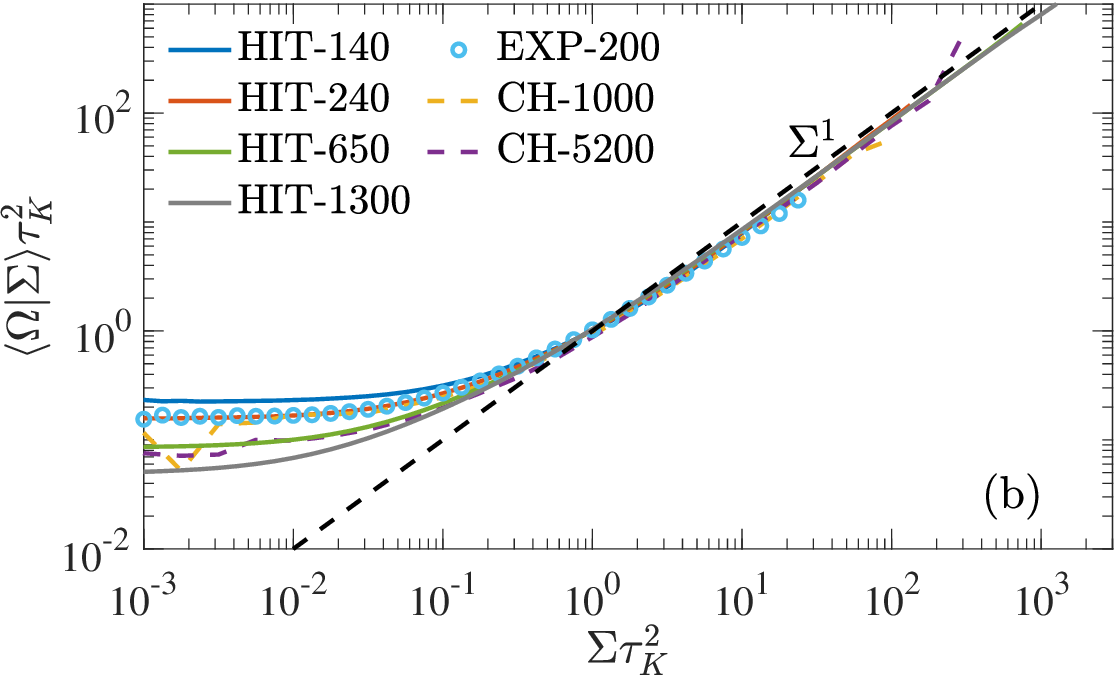}
\includegraphics[width=0.32\textwidth]{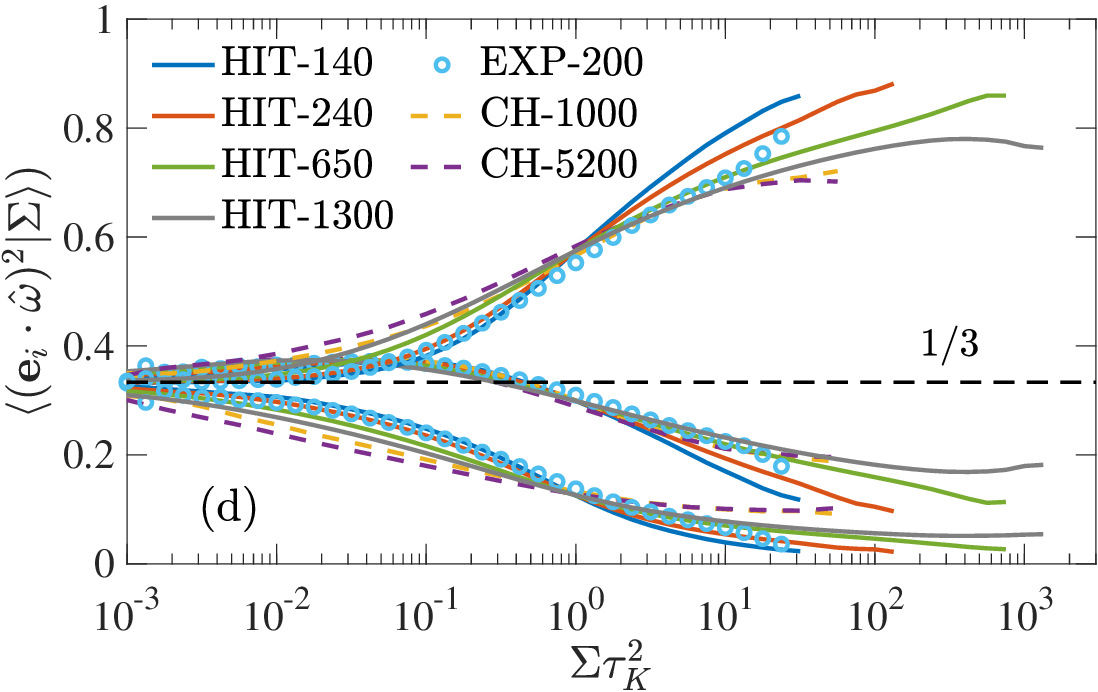}
\includegraphics[width=0.32\textwidth]{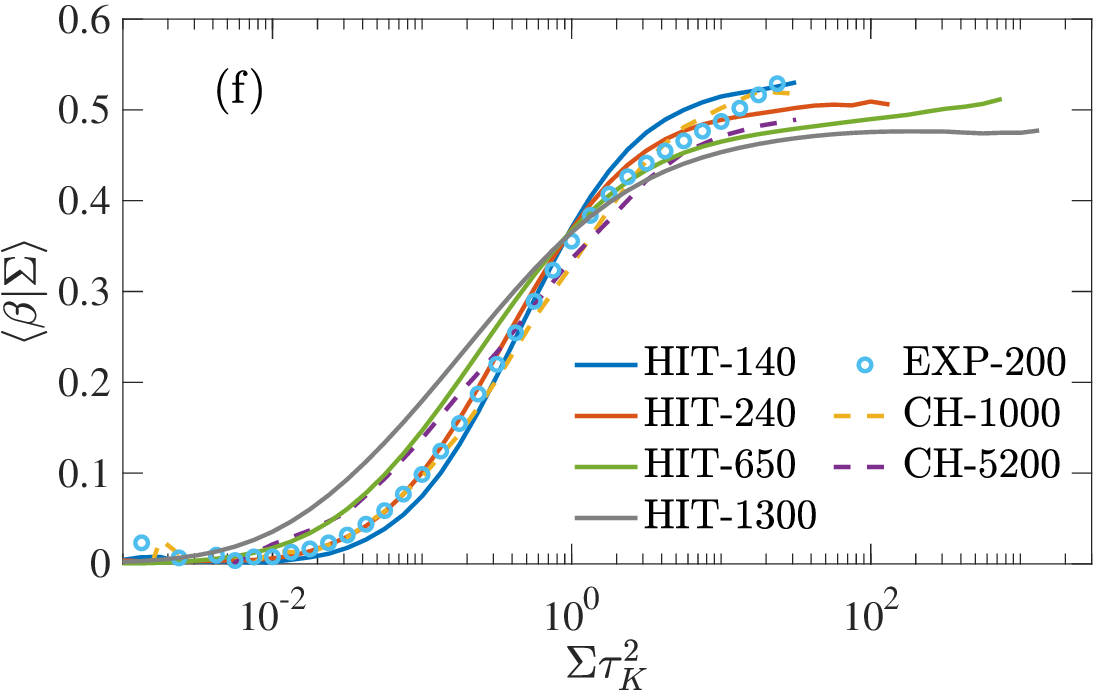} 
\caption{
Conditional expectations of (a) $\Sigma$,  (c) $(\mathbf{e}_i \cdot \hat{\ww})^2$, 
and (e) $\beta$ (e) on $\Omega$, and
of (b) $\Omega$,  (d) $(\mathbf{e}_i \cdot \hat{\ww})^2$, and (f) $\beta$ on $\Sigma$, 
for different flows. See Table~\ref{tab:runs} for description of the runs.
In panel a, the inset shows a zoomed in version of the plot. 
}
\label{fig:cexp}
\end{figure*}

The results in Fig.~\ref{fig:mom2} suggests that
the behavior of velocity gradients in HIT
can be quantitatively extended to other flows
(albeit in regions of no mean shear).
However, one has to be mindful about matching
the appropriate Reynolds numbers. Instead
of simply matching the large-scale or Taylor-scale
Reynolds, one has to match small-scale quantities
such as skewness or flatness of velocity gradients. 
Based on this, we can readily see that 
the channel flow runs at $Re_\tau =1000$ and $5200$
correspond to HIT runs at approximately
$\re=240$ and $650$, respectively, 
substantially larger than the $\re$ directly defined; whereas
the $\re=200$ from von Karman flow is slightly larger than $\re=140$ in HIT.
(Likewise, some minor shifting is required for grid turbulence data 
\cite{gylfason2004} and HIT data from \cite{Ishihara07}).
With these considerations, we next investigate the
structure and geometry of velocity gradient tensor,
providing even stronger support for universality.

To gain deeper insight into extreme gradients,
we separately analyze regions of intense vorticity
and strain \cite{BBP2020, BP2022, BPB2022}, by conditioning
flow properties on: 
\begin{align}
\Omega = \omega_i \omega_i \ , \ \  \
\Sigma = 2 S_{ij} S_{ij} \ , 
\end{align}
where $\Omega$ is the enstrophy
and $\Sigma = \epsilon/\nu$.
In homogeneous flows, $\langle \Omega \rangle = \langle \Sigma \rangle
= 1/\tau_K^2$; thus the extremeness of an event can be quantified by
deviation of $\Omega \tau_K^2 $ (or $ \Sigma  \tau_K^2 $) from unity.
This has allowed us to analyze
the structure of extreme events in HIT
\cite{BPBY2019, BBP2020, BP2022, BPB2022, BP2023}. 
In the following, we compare
key findings
obtained from HIT
with those from other flows.

The first comparison is performed for 
the conditional expectations: 
$\langle \Sigma | \Omega \rangle $ and 
$\langle \Omega | \Sigma \rangle $, which quantity 
the relative magnitudes of strain and vorticity
in regions of intense vorticity and strain, respectively.
In previous works, we showed that for HIT,
$\langle \Sigma | \Omega \rangle \sim \Omega^\gamma $
(for $\Omega \tau_K^2 > 1$),
where $\gamma < 1$ and grows slowly with $\re$; 
additionally, the exponent $\gamma$ can further be related to scaling of 
extreme events and smallest scales of turbulence 
\cite{BPBY2019, BP2022}. 
Figure~\ref{fig:cexp}a shows  
$\langle \Sigma | \Omega \rangle $ for all flows.
We remarkably observe
that the two curves for channel flow, at 
$Re_\tau=1000$ and $5200$, are essentially
identical to HIT curves for $\re=240$ and $650$, respectively -- 
in perfect agreement with earlier inferences 
from Fig.~\ref{fig:mom2}.
This reiterates that extreme events are universal and 
depend only the turbulence intensity.
The result for experiments is also in good agreement
with HIT data, though there are some systematic deviations
for the most extreme events, likely attributable to
uncertainties associated with measuring them.
On the other hand, it is known 
for HIT that $\langle \Omega | \Sigma \rangle \sim \Sigma^1$ 
\cite{BPB2022, BP2022}.
Figure~\ref{fig:cexp}b shows  
$\langle  \Omega | \Sigma \rangle $ for all flows,
and it can be seen that all curves collapse
with a universal $\Sigma^1$ scaling.

We next assess the universality of the structure of gradient tensor,
by considering (as in Fig.~\ref{fig:mean}) 
vorticity-strain alignments
and the quantity $\beta$.
Figure~\ref{fig:cexp}c-d shows the second moments
of alignment cosines, conditioned on $\Omega$ and $\Sigma$ respectively.
The second moment has the useful property:
$\sum_{i=1}^3 (\mathbf{e}_i \cdot \hat{\omega})^2 = 1 $, 
allowing systematic characterization as a function
of conditioning variable and also $\re$-dependence \cite{BBP2020, BPB2022}.
Consistent with earlier results from HIT \cite{BBP2020, BPB2022}, 
Fig.~\ref{fig:cexp}c-d, revealing excellent agreement between
all flows, demonstrates that 
vorticity aligns even strongly with $\mathbf{e}_2$ for extreme events.
Additionally, all alignments are effectively $\re$-independent in regions
of intense vorticity, and exhibit a weak $\re$-dependence in regions
of intense strain.

Figure~\ref{fig:cexp}e-f shows the expectations
of $\beta$, conditioned on $\Omega$ and $\Sigma$ respectively.
Once again, we obtain remarkable agreement between all flows,
in support of universality 
\footnote{There are some minor deviations
for the experimental data, which can be attributed to 
difficulties in resolving large $\Omega$ in experiments}.
The results 
effectively $\re$-independent in regions
of intense vorticity, and exhibit a weak $\re$-dependence in regions
of intense strain. Moreover, $\beta$ is
nearly constant in regions of intense strain, and also larger
than its value in regions of intense vorticity, where
it slowly decreases (with $\Omega$) -- this behavior is
readily explained by predominance of sheet-like
structures for intense strain, and tube-like 
structures for intense vorticity \cite{moisy:2004, BPB2020, BBP2020}. 
Additional results on conditional
vortex stretching and strain self-amplification
are shown in Appendix C and also strongly support universality.


\paragraph{Discussion:}
The K41 phenomenology posits universality of small scales, 
quantitatively characterized by the mean dissipation-rate. 
However, it well known that dissipation-rate and velocity gradients
in general exhibit extreme fluctuations that 
invalidate K41's mean-field description. 
We investigated the universality of extreme events by 
comparing their statistical properties across different turbulent flows,
viz. isotropic turbulence, plane channel flow, and experiments in a von Karman 
mixing tank, also including 
some previous data from DNS \cite{Ishihara07} and  
grid-turbulence experiments \cite{gylfason2004}. 
Our first result is that the scaling exponents 
of velocity gradient moments, as a function of Reynolds number,
are universal across different flows, extending
the findings of \cite{Schumacher+14} to higher 
Reynolds numbers. 
Using an approach in the spirit of 
extended self-similarity~\cite{Benzi+93},
we further demonstrate that even the proportionality constants 
are universal, provided the gradient moments are 
appropriately matched across flows, which occurs at different
Reynolds numbers in different flows.

Thereafter, we present a detailed
comparison of the structure of the velocity gradient
tensor in different flows, by considering
various statistics conditioned on extreme
vorticity and strain. Once again, the 
results from isotropic turbulence
are in near perfect quantitative agreement
with those in other flows, once the Reynolds
numbers are matched using prior results
from gradient moments. 
Overall, our results reinforce
small-scale universality
for mean and extreme events alike. 
While K41's mean-field description
needs to replaced by intermittency models,
our works show that results from isotropic turbulence can 
accurately characterize extreme events in other flows, extending 
well beyond the scope of scaling exponents alone. 
It also suggests that velocity gradients would serve as a more
effective modeling target for capturing small-scale dynamics; 
such an approach could be particularly advantageous when 
leveraging machine learning techniques \cite{tian21, bs_pnas_2023}, 
since they rely on directly learning from DNS data which are restricted
to lower Reynolds. 

Finally, it is worth noting that the observed
universality in different flows
is obtained far from boundaries
and in regions of weak to no mean-shear (i.e., large-scale gradient).
Given the evidence from other flows \cite{Pumir:1995,Shen:2000, BifProc} 
and also scalar turbulence \cite{Sreeni91, Pumir94, Warhaft00,BCSY2021a},
it appears that the presence of a large-scale
mean-gradient persistently disrupt local isotropy and 
universality, even at high Reynolds numbers \cite{Shen:2000}.  
More effort is necessary to understand
this effect in detail, and additionally  develop
intermittency theories capable of incorporating the effects of 
large-scale mean-gradients.

\begin{acknowledgements}
\paragraph*{Acknowledgments:}
We gratefully acknowledge the Gauss Centre for Supercomputing 
e.V. (www.gauss-center.eu) for providing computing time on the supercomputers 
JUQUEEN and JUWELS at J\"ulich Supercomputing Centre (JSC),
where the simulations reported in this
paper were performed. We are very grateful to
Eberhard Bodenschatz and Gerhard Nolte for making the 
experimental data
available for processing, and also thank Armann Gylfason and Zellman Warhaft
for sharing previous results, used in Figs.2-3.  
\end{acknowledgements}



%


\subsection*{Appendix A -- Methods}

\paragraph*{DNS data:}  
The DNS data were obtained by solving the 
incompressible Navier-Stokes equations:
\begin{equation}
\partial \uu / \partial t + ( \uu \cdot \nabla) \uu = - \nabla p + \nu \nabla^2 \uu + \mathbf{f} \label{eq:NS}  \, ,
\end{equation}
where $\uu$ is the divergence-free velocity $(\nabla \cdot \uu = 0)$, $p$ is the 
kinematic pressure, 
and $\mathbf{f}$ is a large-scale forcing term, 
which depends on the flow considered.

The first flow considered is the canonical setup of forced
isotropic turbulence with periodic boundary conditions,
which allows us to use efficient and
highly accurate Fourier pseudo-spectral methods \cite{Rogallo};
The largest scales are forced isotropically to maintain a 
statistically stationary state \cite{EP88}.  
The domain size is $(2 \pi)^3$, discretized into $N^3$ grid points, 
with uniform grid-spacing $\Delta x = 2\pi/N$ in each direction. 
The Taylor-scale Reynolds number $\re$ goes from $140$ to $1300$ -- 
as listed in Table~\ref{tab:runs}, and special attention is given to resolve
the small scales and extreme events \cite{BPBY2019}, with $\Delta x/\eta_K \sim 0.5$. 
Additional details about the database can be found in several recent 
studies \cite{BBP2020, BPB2020, BS2020, BP2021, BPB2022, BS2022, BP2023},
which have also adequately established convergence
with respect to resolution and statistical sampling.

The second flow considered is plane channel flow between two parallel plates
(wall normal coordinate being $y$), 
forced by a constant mean pressure gradient. 
This data is obtained 
from the Johns Hopkins turbulence database \cite{Graham+15},
and corresponds to skin-friction Reynolds numbers $Re_\tau=1000$ and $5200$.
Note, $Re_\tau = u_\tau \delta/\nu$, where $u_\tau$ is the skin
friction velocity and $\delta$ is the half-width of channel.
For both cases, the domain size is $8\pi \delta \times 2\delta \times 3\pi\delta$, 
however, the grid spacing is not the same in each direction.
Table~\ref{tab:runs} lists the grid spacing  $\Delta x_\alpha/\eta_K$
in each coordinate direction, 
at the center of the channel. 
It is worth noting that the resolution in the streamwise 
direction is $\Delta x_1/\eta \approx 2.2$ for $Re_\tau=1000$, and $1.5$ for 
$Re_\tau=5200$.  Earlier results from HIT \cite{BPBY2019} suggest that given the turbulence
intensity of these flows, the resolution is not fully sufficient to capture the 
extreme events of velocity gradients (the component $A_{11}$ in this case). 
As discussed in Appendix B, this also appears to be the reason why 
the high order moments of $A_{11}$ are somewhat unpredicted.

\paragraph*{Experimental data:}
The experimental data were obtained at the center of a von Karman mixer, 
where the fluid is set to motion by two counter rotating impellers rotating 
at $0.2 {\rm Hz}$, with the axis of rotation assumed to be in the $y$ direction.
The Taylor-scale Reynolds number is about $\re = 200$.
The apparatus and the data acquisition method are also 
described in~\cite{Knutsen:2020,BLW:2024}. 
We note that the 
scanning Particle Image Velocimetry (PIV)
technique allows us to obtain the full velocity field (and hence also
the gradient field) in an observation volume of size $(42 \eta_K)^3$, 
with $\eta_K \approx 240 \mu m$,  over a measurement grid-spacing
of $\Delta x \approx 0.8 \eta_K$. Thus, each snapshot corresponds
to about $53^3$ samples; about $10^5$ such snapshots of the flow
where analyzed to obtain the desired statistics.

\subsection*{Appendix B -- Local isotropy of velocity gradients}

Small-scale isotropy or local isotropy is a prerequisite
for universality \cite{K41a, Frisch95}.
For the velocity gradient tensor, it can assessed 
rigorously by considering various moments of all nine components \cite{popebook},
but to keep things straightforward, we will focus on longitudinal (diagonal) components, 
$A_{\alpha \alpha}$ (no summation implied) for $\alpha=1,2,3$. 
For local isotropy to be satisfied,  one expects
the moments $ M_n = \langle A_{\alpha \alpha}^n \rangle \tau_K^n $ 
to be identical  for $\alpha=1,2,3$.
This is indeed the case for HIT, and thus, we report
only a single value in Table~\ref{tab:runs} for all HIT runs. 
However, for channel flow and experiments, 
we report three numbers for $\alpha=1,2,3$,
to assess the validity of local isotropy.

As mentioned in the the main text, for $n=2$,
local isotropy dictates 
$\langle \epsilon \rangle = 15 \nu  A_{\alpha \alpha}^2$
and thus $15 M_2 = 1 $.
We observe in Table~\ref{tab:runs} that this is indeed
the case, with some minor deviations for non-HIT runs, 
especially in experiments. 
For the skewness, given by $\mathcal{S}=M_3/M_2^{3/2}$,
we observe more noticeable departures from isotropy
for both channel flow and experiments, suggesting
some effect of anisotropic large-scale forcing.
Note that the skewness is different in $y$-direction
for both channel flow and experiments, consistent with direction
of large-scale anisotropy.
For channel flow, this effect seems to be diminishing 
with increasing $Re_\tau$, suggesting local isotropy
would be strictly recovered at higher Reynolds number.
The departure is more noticeable for experiments, likely
because of presence of a weak mean-strain at the center of 
the tank \cite{Knutsen:2020}.

Since the skewnesses for longitudinal components are not equal,
we define 
$ \mathcal{S}^* =  (6\sqrt{15}/7)\langle \omega_i \omega_j S_{ij} \rangle \tau_K^3$,
which measures vortex stretching 
and serves as an effective skewness; the rationale being that for local
isotropy $ \mathcal{S}^*  = \mathcal{S}$ \cite{BBP2020}. 
Alternatively, one can also use the quantity
$\langle S_{ij} S_{jk} S_{ki} \rangle$, which measures
self-amplification of strain-rate \cite{Tsi2009}.
In homogeneous turbulence, 
$\langle S_{ij} S_{jk} S_{ki} \rangle = - 
\frac{3}{4} \langle \omega_i \omega_j S_{ij} \rangle$ \cite{Betchov56},
and although not shown, this relation is near-perfectly satisfied
for all the flows considered here.

\begin{table*}[!hbtp]
    \begin{tabular}{l|c|c|c|c|c||c|c|c}
    \hline
    case      & HIT-140 & HIT-240 & HIT-390 & HIT-650 &  HIT-1300 & CH-1000 & CH-5200 & EXP-200 \\
    \hline
    $Re_\tau$ & -   & -  & - & -   & -     & 1000   & 5200    & -  \\
    $Re_\lambda$ & 140   & 240  & 390 & 650   & 1300     & 63   & 156    & 200  \\
    $\Delta x_\alpha / \eta_K$         & 0.5  & 0.5   & 0.5   & 0.5  & 0.5   & 2.2, 1.1, 1.1    & 1.5, 1.2, 0.8    & 0.8   \\
    $15 M_2 $         & 1.0  & 1.0   & 1.0   & 1.0  & 1.0   & 0.96, 1.03, 1.00    & 0.99, 1.01, 1.00    & 1.10, 0.90, 1.10   \\
    $-\mathcal{S} = - M_3/M_2^{3/2}$        & 0.524   & 0.552   & 0.585   & 0.630  & 0.695    & 0.439, 0.699, 0.468      & 0.579, 0.724, 0.589   &  0.750, 0.100, 0.775   \\    
    $-\mathcal{S}^*$       & 0.524   & 0.56   & 0.585  & 0.630  & 0.695   & 0.57    & 0.64    & 0.524   \\
    $\mathcal{F} = M_4 / M_2^2$         & 5.74   & 6.82   & 8.02  & 9.90  & 13.1   & 6.44, 7.35, 6.80    & 9.70, 10.3, 10.0     & 6.2, 6.3, 6.2   \\
    $\mathcal{H} = M_6/M_2^3$         & 113   & 200   & 354  & 609  & 1495   & 134, 207, 161    & 472, 601,  626     & 122, 130, 118   \\
    \hline
    \end{tabular}
\caption{
Simulation parameters and velocity gradient moments for various
flows investigated. Cases `HIT-' and `CH-' respectively correspond to 
HIT and channel flow DNS, and `EXP' corresponds to experiments.
$\Delta x_\alpha/\eta_K$ is the grid spacing in DNS and measurement
resolution in experiments. 
Moments $M_n$ are non-dimensional moments of longitudinal
velocity gradients $A_{ \alpha \alpha }$, as defined by Eq.~\eqref{eq:mn}. 
The effective skewness is defined as 
$ \mathcal{S}^* =  (6\sqrt{15}/7)\langle \omega_i \omega_j S_{ij} \rangle \tau_K^3$ 
\cite{BBP2020}. 
For HIT, only one set of values are reported, since the values in all three coordinate
directions are essentially identical (within $2\%$ of each other or less), 
but channel flow and experiments exhibit
some weak anisotropy. 
}
\label{tab:runs}
\end{table*}

Finally, we consider flatness $\mathcal{F} = M_4/M_2^2$ 
and hyper-flatness $\mathcal{H} = M_6/M_2^3$ of the 
gradients. Same as for $M_2$, we observe that local
isotropy is reasonably satisfied for channel flow and experiments.
However, for channel flow, the hyper-flatness for $A_{11}$ is
noticeably lower. We believe this can be attributed
to lack of small-scale resolution in the $x$-direction. 
It can be seen from the table, that $\Delta x_1 /\eta_K $
for channel flow does not resolve adequately resolve
the Kolmogorov length scale \cite{BPBY2019}. 
For this reason, when considering the fourth moments,
we average over all three directions,
but for sixth moments for 
channel flow, we simply take the average
over $y$ and $z$ directions for a reliable measure.
Given the restrictions on resolution and statistics
for non-HIT flows, 
we refrain from considering moments higher than $n=6$
in this work. \\

\begin{figure}
\centering
\includegraphics[width=0.43\textwidth]{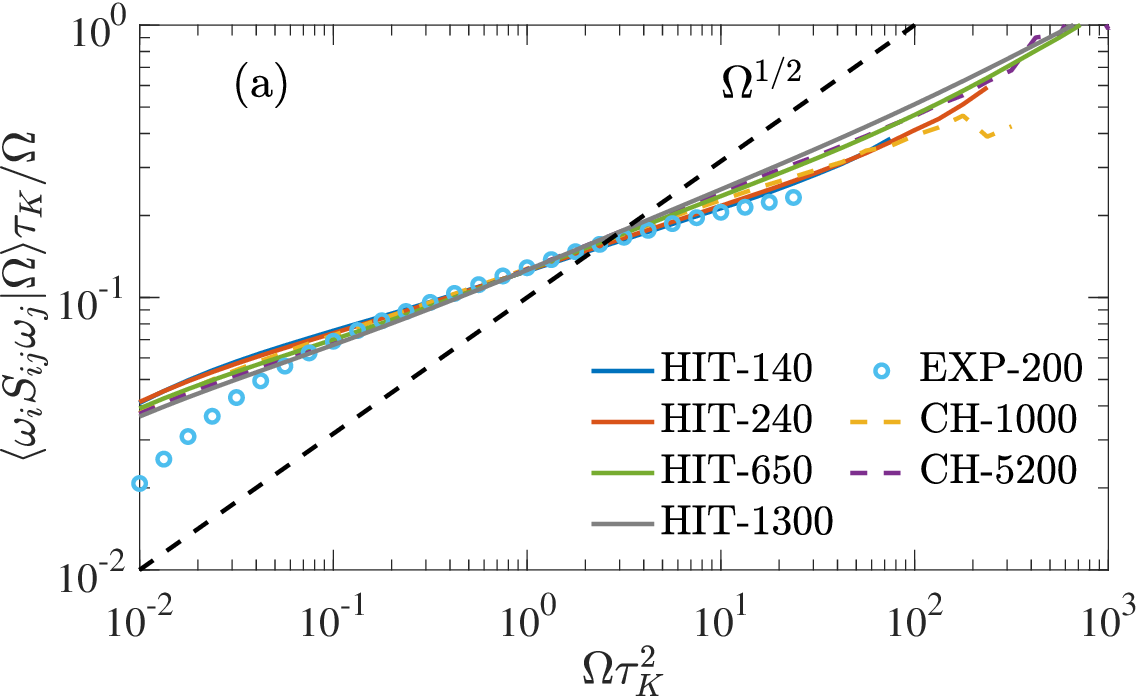} 
\includegraphics[width=0.43\textwidth]{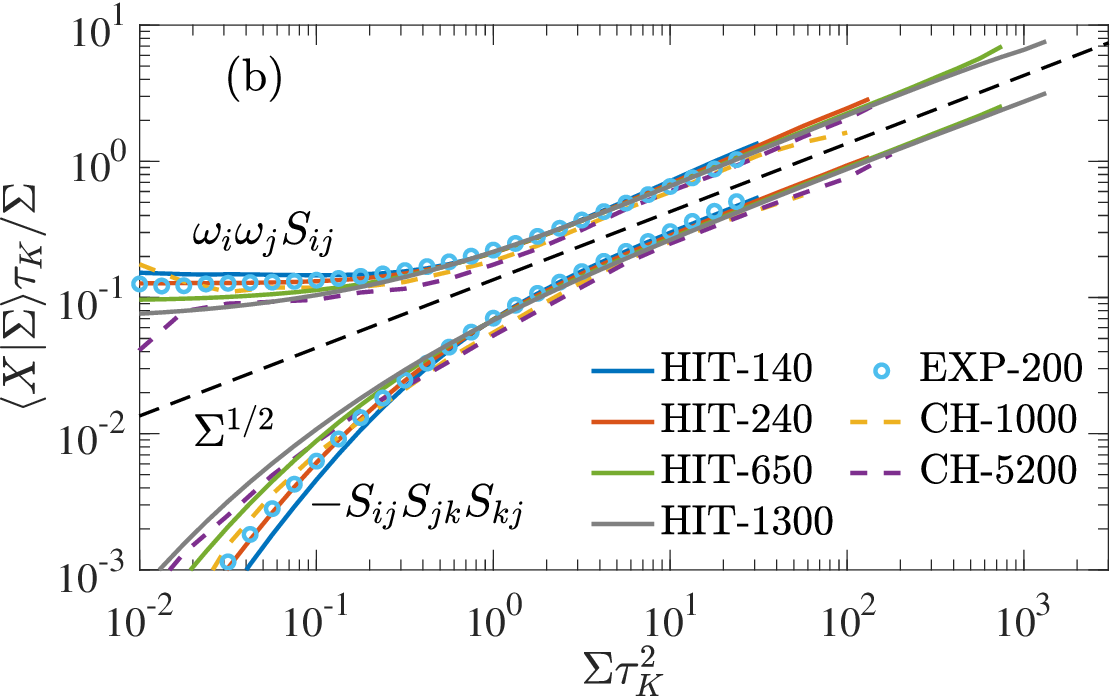} 
\caption{
(a) Expectation of vortex stretching
$\omega_i \omega_j S_{ij}$, conditioned on $\Omega$.
(b) Expectations of vortex stretching, $\omega_i \omega_j S_{ij}$,
and strain self-amplification,  
$ S_{ij} S_{jk} S_{kj}$, conditioned on $\Sigma$.
In both plots, the black dashed lines correspond to 
power-law of $\text{slope}=1/2$. 
}
\label{fig:ws}
\end{figure}

\subsection*{Appendix C -- Universality of vortex stretching and 
strain self-amplification}

To further add to the results presented in Fig.~\ref{fig:cexp}, 
we consider here the conditional expectations of the
vortex stretching, $\omega_i \omega_j S_{ij}$,
and strain self-amplification,  
$ S_{ij} S_{jk} S_{kj}$, which 
appear in the transport equations for $\Omega$ and $\Sigma$
\cite{Tsi2009, BBP2020, BPB2022}. 
Figure~\ref{fig:ws}a shows 
$\langle \omega_i \omega_j S_{ij} | \Omega \rangle $
for all flows, whereas Fig.~\ref{fig:ws}b shows
both terms conditioned on $\Sigma$.
Note that only vortex stretching is shown in
Fig.~\ref{fig:ws}a, since  strain self-amplification does
not contribute to vorticity amplification (but vortex stretching
does play role in strain amplification).

A detailed discussion of the 
results in Fig.~\ref{fig:ws}a-b
and their implications, 
can be found in previous works \cite{BBP2020, BPB2022};
but it is worth noting that the qualitative
behavior of the curves in Fig.~\ref{fig:ws}a-b is quite 
similar to those in Fig.~\ref{fig:cexp}a-b, respectively.
Essentially, the quantity 
$\langle \omega_i \omega_j S_{ij} | \Omega \rangle  \tau_K /\Omega$ 
grows as $\Omega^\beta$ for extreme events, where $\beta < 1/2$ as expected
from a simple scaling relation between strain and vorticity. 
Moreover, the exponent $\beta$ weakly increases 
with Reynolds, slowly approaching the limiting value of $1/2$.
On the other hand, the quantities 
$\langle \omega_i \omega_j S_{ij} | \Sigma \rangle  \tau_K /\Sigma$ 
and 
$-\langle S_{ij} S_{jk} S_{kj} | \Sigma \rangle  \tau_K /\Sigma$ 
both grow as $\Sigma^{1/2}$ for extreme events, with nearly no 
dependence on Reynolds number. 

However, the key new observation is the agreement
of results from different turbulent flows.
Similar to previous result in 
Fig.~\ref{fig:cexp}a, we observe in
Fig.~\ref{fig:ws}b that the results for
HIT at $\re=240$ and $650$ are essentially identical to 
those from channel flow at $Re_\tau=1000$ and $5200$, respectively.
In conclusion, the results in 
Fig.~\ref{fig:ws}a-b once again strongly reinforce the 
universality of extreme events across different flows.

\end{document}